%% file: globecom2019.tex
\documentclass[conference]{IEEEtran}
\IEEEoverridecommandlockouts
\usepackage{cite}
\usepackage{graphicx}
\usepackage{subcaption}
\usepackage{xcolor}
\usepackage{tikz}
\usepackage{pgfplots}
\pgfplotsset{compat=newest}
\usepgfplotslibrary{groupplots}
\usepackage{multirow}
\usepackage{upgreek}
\usepackage{amssymb}
\usepackage{amsmath}
\usepackage{cases}
\usepackage{pifont}
\usepackage{balance}
\usepackage{bm}
\usepackage{amsthm}
\usepackage{tabularx}
\usepackage{booktabs}
\newcolumntype{Y}{>{\centering\arraybackslash}X}
\usepackage{array}
\usetikzlibrary{arrows,positioning,shapes.geometric,spy, matrix}
\usepackage{float}
\usepackage{changes}
\usepackage{bbm}
\def\BibTeX{{\rm B\kern-.05em{\sc i\kern-.025em b}\kern-.08em
		T\kern-.1667em\lower.7ex\hbox{E}\kern-.125emX}}
\newcommand{\bfu}{\mbox{\boldmath $u$}}
\newcommand{\bfx}{\mbox{\boldmath $x$}}
\newcommand{\bfy}{\mbox{\boldmath $y$}}
\newcommand{\bfp}{\mbox{\boldmath $p$}}
\newcommand{\bfs}{\mbox{\boldmath $s$}}
\newcommand{\bfr}{\mbox{\boldmath $r$}}
\newcommand{\bfn}{\mbox{\boldmath $n$}}

\newcommand{\fixme}[2]{\ifx&#2&{\leavevmode\color{red}#1}\else{\{}\textcolor{red}{#1}{\}}\footnote{{\leavevmode\color{red}#2}}\PackageWarning{Fixme}{#1: #2}\fi}

\begin{document}

	\title{Efficient Flicker-Free FEC Codes using Knuth's Balancing Algorithm for VLC}
	
	\author{\IEEEauthorblockN{Elie Ngomseu Mambou\IEEEauthorrefmark{1}, Thibaud Tonnellier\IEEEauthorrefmark{1}, Seyyed Ali Hashemi\IEEEauthorrefmark{2}, and Warren J. Gross\IEEEauthorrefmark{1}}
		\IEEEauthorblockA{\IEEEauthorrefmark{1}Department of Electrical and Computer Engineering, McGill University, Montr\'eal, Qu\'ebec, Canada\\
			\IEEEauthorrefmark{2}Department of Electrical Engineering, Stanford University, Stanford, California, USA\\
			elie.ngomseumambou@mail.mcgill.ca, thibaud.tonnellier@mail.mcgill.ca, ahashemi@stanford.edu, warren.gross@mcgill.ca}}
	
	\maketitle
	
	\begin{abstract}
		Visible light communication (VLC) provides a short-range optical wireless communication through light-emitting diode (LED) lighting. Light beam flickering and dimming are among the challenges to be addressed in VLC. Conventional methods for generating flicker-free codes in VLC are based on run-length limited codes that have poor error correction performance, use lookup tables which are memory consuming, and have low transmission rates.
		In this paper, we propose an efficient construction of flicker-free forward error correction codes to tackle the issue of flickering in VLC. Our simulation results show that by using polar codes and at a dimming ratio of $50\%$, the proposed system generates flicker-free codes without using lookup tables, while having lower complexity and higher transmission rates than the standard VLC methods.
		For an information block length of $256$, the error correction performance of the proposed scheme is $1.8$ dB and $0.9$ dB better than that of the regular schemes at the bit error rate of $10^{-6}$ for a rate of $0.44$ and $0.23$, respectively.
	\end{abstract}
	
	\begin{IEEEkeywords}
		Visible light communication, run-length limited, flicker-free, polar codes, successive cancellation decoding. 
	\end{IEEEkeywords}
	
	
	\section{Introduction} \label{s1}
	The number of global mobile device subscribers is estimated at $5.5$~billion by 2022. Due to this exponential growth, alternative methods of communication should be investigated to decongest the RF bandwidth. Visible light communication (VLC) could be considered in further communication standards to handle the short range optical wireless communication (OWC) through light beam \cite{gupta2015}. 
	Advantages of VLC technology include, but are not limited to, low cost deployment, higher security, lower interference from RF devices, and $10000$ times bigger unregulated bandwidth compared to RF systems \cite{haas2013}. VLC covers a broad spectrum range from $350$~THz to $750$~THz. An overview and an introduction on VLC can be found in \cite{pathak2015}.
	
	The VLC field requires efficient modulations and forward error correction (FEC) codes to address challenges such as flickering and uncontrolled light dimming. In the objective of solving these issues, some standards have been defined. On-off keying (OOK), variable pulse-position modulation (VPPM), and orthogonal frequency division multiplexing (OFDM) are considered in modulation, while Manchester codes or 1b2b, 4b6b, and 8b10b are chosen as run-length limited (RLL) coding based on Reed-Solomon and convolutional codes \cite{ieee2011st}. However, the soft-output decoding of RLL codes requires a high computational complexity.
	
	Various FEC coding schemes have been described to improve the proposed standards based on Reed-Muller (RM) codes and compensation symbols (CS) \cite{kim2013}, low-density parity-check (LDPC) codes \cite{kim2015}, and turbo codes \cite{lee2012}. More recently, polar codes have been introduced in VLC \cite{fang2017, wang2017}.
	
	
	Since the encoding of polar codes does not generate flicker-free codewords, some constraints are imposed to control the light variation. In \cite{fang2017}, a fixed number of CS is appended at the tail of the polar codeword according to the dimming ratio and using OOK modulation. In \cite{yao2017}, a modification of the traditional polar code kernel was presented to improve the bit error rate (BER) for VLC systems and decrease the complexity of hardware implementation. A modified likelihood ratio (LR) of polar code decoding for inter-symbol interference (ISI) is proposed in \cite{wang2017} which improves the conventional LR function for ISI in VLC channel. A design of polar codes for RGBA (red-green-blue added with amber chip to control white light) light-emitting diode (LED) channel in OOK is presented in \cite{jiang2017}; results show that the added amber chip improves the error correction performance under the successive cancellation (SC) list decoding of polar codes for red-green-blue (RGB) LED channel. Although the aforementioned methods improve the error correction performance, they present a low data transmission rate and make use of variable CS length which increases the decoding complexity. In addition, some of these methods still use RLL coding.
	
	In this paper, we propose an efficient construction of flicker-free FEC codes based on Knuth's balancing algorithm \cite{knuth1986}. Constrained sequence coding has found applications in several other fields such as magnetic and optical recording devices, detection and correction of unidirectional errors, cable transmissions, and noise attenuation in VLSI systems \cite{immink2004, mambou2017j}. The proposed coding scheme does not make use of look-up tables which are memory-consuming, and the low redundancy provided by Knuth's balancing algorithm is exploited for higher data transmission rates compared to RLL based schemes while providing less complexity.
	
	
	\section{Preliminaries} \label{s2}
	
	\subsection{VLC Basics}\label{s2.1}
	VLC is a subset of OWC that handles short-range transmission through lighting. That is a data transmission and reception medium which uses the visible light of frequencies between $350$~THz and $750$~THz (wavelengths between $380$~nm and $780$~nm). LEDs are mostly used in VLC as their current intensity is easily modulated and they can transmit signals at around $500$~Mbit/s. In low data rate communications, an Ethernet speed of $10$~Mbit/s can be achieved at a distance between $1$ to $2$ kilometres \cite{richard2015}.
	
	There are several modulation techniques used in VLC. The most used are: OOK, digital representation of light as $1$ or $0$ corresponding to the presence or absence of information; pulse position modulation (PPM), encoding $m$ message bits by sending a single pulse in one of $2^m$ possible time-shifts; frequency shift keying (FSK), transmission of information through discrete frequency changes of a carrier wave; and colour shift keying (CSK), mapping data from stream symbols to colours that are produced by the tristimulus principle based on the RGB system.
	
	The brightness control of the LED is a relevant challenge in all the aforementioned modulation schemes. To avoid flickering, a lighting spectrum must be encoded into a flicker-free binary stream depending on the optical clock rate. The maximum flickering time period (MFTP) refers to the maximum time period that is flexible to light intensity variations and not perceived by human eye. A frequency of at least $200$~Hz is required to mitigate flickering \cite{ieee2015}, which corresponds to a MFTP of shorter than $5$~ms.
	
	For the scope of this work, we will be using the binary phase-shift keying (BPSK) modulation through an additive white Gaussian noise (AWGN) channel for simplicity purposes. However, the proposed coding scheme is universal in the sense that it can be used with other modulation techniques and other communication models.
	
	\subsection{Polar Codes}\label{s2.2}
	
	
	
	The mechanism from which polar codes are named is called channel polarization \cite{arikan2009}. Via the channel polarization, two copies of a channel $W$ are transformed into two synthetic channels, such that one becomes stochastically upgraded and the other becomes stochastically degraded. The generator matrix $\bm{G}$ for a polar code of length $N$ is obtained by computing the $n\textsuperscript{th}$ Kronecker power, denoted $\otimes$, of $\bm{F_2}$: 
	$\bm{G} = \bm{F_2}^{\otimes n}$, where $n = \log_2N$ and $\bm{F_2} = \big[ \begin{smallmatrix} 1\ 0 \\ 1\ 1 \end{smallmatrix} \big]$.
	
	A polar code encoding $K$ message bits into a codeword of size $N$ is denoted by $\mathcal{PC}(N,K)$ and its code rate is $R = \frac{K}{N}$. To encode a $K$-sized message, it is necessary to extend its size to $N$ since $\bm{G}$ is a $N \times N$ matrix. To do this, $N-K$ indices, called frozen bits, are inserted into the message. The information bits are
	placed in the $K$ most reliable synthetic channel indices, while the frozen bits are placed on the remaining locations. 
	Once the message is extended into $\bm{u}$, encoding is performed via the matrix multiplication: $\bm{x} = \bm{u}\bm{G}$.
	
	To determine the information bits locations $\mathcal{I}$ and the frozen bits locations $\mathcal{F}$, the reliabilities of each synthetic channel have to be ranked. For AWGN channels, density evolution or Gaussian approximation (GA) can be used \cite{Trifonov2012}. For achieving any length that is not a power of two, different techniques have been proposed \cite{Tonnellier2019}.
	
	The original decoder used to decode polar codes is known as SC decoder. Proposed by Ar{\i}kan in \cite{arikan2009}, it was used to prove the capacity-achieving property of polar codes at infinite block length. The SC decoder can be visualized as binary tree traversal with left-branch-first priority. The tree has a depth of $n+1$ and $N$ leaf nodes.
	Therefore, the number of decoding operations is given by $N\log_2 N$.
	
	Using the GA, one can predict the decoding performance of polar codes under SC decoding in terms of frame error rate (FER) \cite{wu2014}: 
	\begin{equation}
	FER=1-\prod_{i\in \mathcal{I}}\Big(1-\bm{Q}_i\Big)
	\end{equation}
	where the $\bm{Q}$ function is such that $\bm{Q}_i=\frac{1}{2}\mbox{erfc}(\frac{\sqrt{\bm{c}_i}}{2})$ and $\bm{c}$ is the output of the GA algorithm.
	
	\subsection{RLL codes}
	In order to obtain the best decoding performance of a serially concatenated scheme considering an outer FEC and an inner RLL code, the RLL decoding has to provide soft outputs. This enables using soft input for the FEC decoding and thus, greatly improve the overall decoding performance. The 1b2b RLL code can be seen as a Manchester code. Therefore, encoding and decoding are quite simple. The 4b6b encoding requires a lookup table \cite{ieee2011st}. Regarding the decoding, since the 4b6b code is not structured, a plain and complex \emph{a posteriori} probability (APP) decoding has to be carried out in order to obtain soft outputs.
	

	\subsection{Knuth's Balancing Algorithm}\label{s2.4}
	The celebrated Knuth's balancing algorithm \cite{knuth1986} is a simple and efficient method to generate balanced codewords by inverting the first $e$ bits of $\bfx$ with $(1\leq e\leq N)$. The index $e$ is encoded as the prefix $\bfp$ and appended to the generated balanced codeword $\bfx'$. The concatenated sequence $\bfx'\cdot\bfp$ is obtained at the receiver, and then $\bfp$ is used to recover the information word $\bfx$. All these words are translated within the bipolar alphabet $\mathbb{A}^2\in \{-1,1\}$. Let $d(\bfx)$ be the disparity of $\bfx$ as $d(\bfx)=\sum_{i=1}^{N}x_i$. $\bfx$ is said to be balanced if and only if $d(\bfx)=0$. Similarly, the \textit{running digital sum} (RDS) over the first $j$ bits of $\bfx$ is denoted as $d_j(\bfx)$, where $d_j(\bfx)=\sum_{i=1}^{j}x_i$ with $1\leq j\leq N$.
	
	Knuth's algorithm can be comprehended as splitting a word into two segments where the first segment has flipped bits while the second segment remains unchanged, therefore, $d(\bfx)=-\sum_{i=1}^{e}x_i+\sum_{i=e+1}^{N}x_i$. This process always leads to a balanced codeword because an index $e$ is always achievable since $d_{j+1}(\bfx)=d_j(\bfx)\pm2$. The redundancy of the full set of balanced codewords equals $H=N-\log_2 \binom{N}{N/2}$. For large $N$, the redundancy of the Knuth's scheme is almost twice larger than $H$.
	
	\begin{figure*}[th]
		\centering
		\includegraphics[width=.95\linewidth]{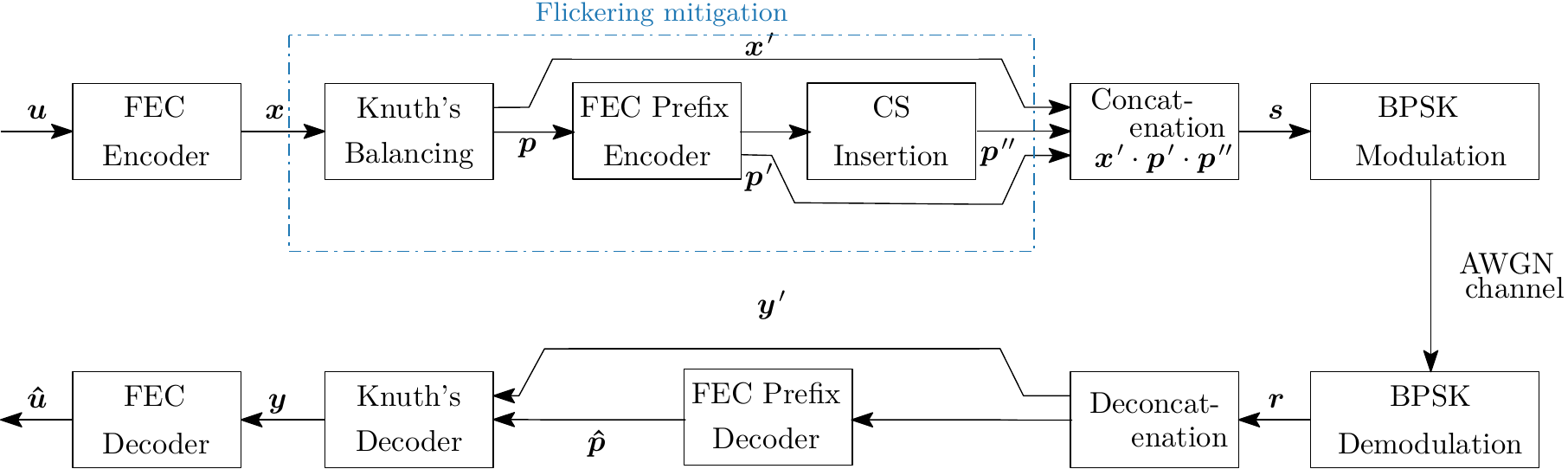}
		\caption{The proposed FEC coding scheme.}
		\label{fig:algo1}
	\end{figure*}
	For example, \eqref{a1} and \eqref{a2} show the balancing of sequences $1011111$ and $100001$ respectively through Knuth's algorithm.
	\begin{align}
	101111&\rightarrow \textbf{0}01111 \rightarrow \textbf{01}1111 \rightarrow\textbf{010}111 \rightarrow \textbf{0100}11.\label{a1}\\
	100001&\rightarrow \textbf{0}00001\rightarrow \textbf{01}0001 \rightarrow \textbf{011}001\label{a2}.
	\end{align}
	The bold symbols represent the inverted segment within the codeword. The balanced state occurs at index $4$ ($100$) in \eqref{a1} and at index $3$ ($011$) in \eqref{a2}. 
	
	\section{Proposed Coding Scheme} \label{s3}
	
	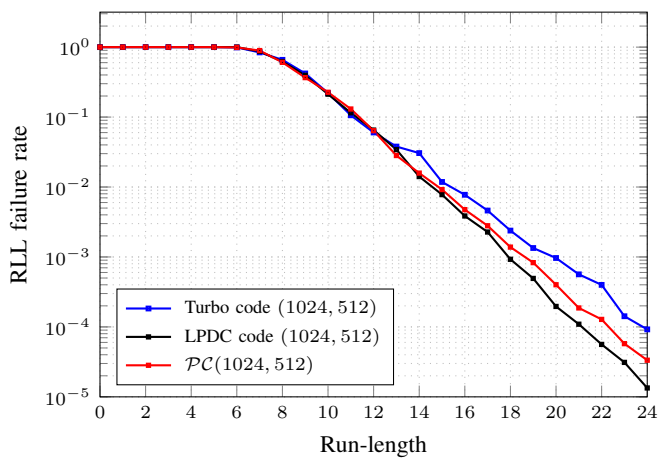
\begin{figure}
		\centering
		\input{img/rllResults.tikz}
		\caption{Run-length performance comparison for various codes with $R=0.5$.}
		\label{rlls}
	\end{figure}
	
	\subsection{The Proposed Scheme}\label{s3.1}
	The block diagram of the proposed coding scheme for VLC system is presented in Fig.~\ref{fig:algo1}. The transmitter is made by a FEC encoder followed by Knuth's balancing algorithm encoder. The generated prefix is  encoded again, using another FEC, and balanced through CS insertion. The resulting flicker-free signal is then transmitted using BPSK modulation. At the receiver, the signal is received and demodulated. Subsequently, the original message can be recovered by undoing the encoder steps in a reverse direction.
	
	The proposed scheme can be used with any FEC code. In Fig.~\ref{rlls}, the flicker-free signal $\bfs$ was estimated distinctly for a Turbo, an LDPC, and a polar code of length $1024$ with $R=0.5$. The test consisted of counting the number of frames 
	that does not match the run-length constraint. At least $100$ frame ``errors'' for each run-length within $l\in\{0,1,\dots,20\}$ were counted. It can be observed that the RLL failure rate decreases with the increase of run-lengths.
	
	In the following, polar codes are used as the underlying FEC code in the proposed scheme. A message $\bfu$ is polar encoded as $\bfx$ of block length $N$. The balancing of $\bfx$ is then performed through Knuth's algorithm resulting in $\bfx'\cdot\bfp$, where $\bfx'$ is the balanced codeword from $\bfx$, and $\bfp$ is the appended prefix of length $p=\log_2N$. This prefix $\bfp$ is polar encoded once again through $\mathcal{PC}(p',p)$. The new encoded prefix $\bfp'$ is then balanced by appending the vector $\bfp''$ which is the $1$'s complement of $\bfp'$. The transmitted codeword $\bfs$ of length $S=N+2p'$ is the concatenation of $\bfx'$, $\bfp'$, and $\bfp''$, $\bfs=\bfx'\cdot\bfp'\cdot\bfp''$ which is modulated using BPSK as shown in Fig.~\ref{fig:algo1}. The transmission rate of the system is $R=\frac{K}{N+2p'}$. The vector $\bfr=\bfs+\bfn$ is received after demodulation, where $\bfn$ designates the Gaussian noise vector. Then, through the decoded prefix, $\bfy$ is retrieved and parsed to the polar decoder as shown in Fig.~\ref{fig:algo1}, and the estimated message is recovered as $\hat{\bfu}$.
	
	For the two codewords, $\bfp'$ and $\bfp''$, the Manchester code is decoded through a soft-input soft-output decision as $p'_i-p''_i$ where $p'_i$ and $p''_i$ are elements of $\bfp'$ and $\bfp''$ respectively with $1\leq i\leq p'$. The resulting  $p'_i-p''_i$ are inputs to the FEC prefix decoder which performs a SC decoding.
	
	\subsection{Redundancy Study}\label{s3.2}
\begin{figure}[t]
		\centering
		\input{img/figure.tikz}
		\caption{Redundancy performance comparison.}
		\label{red}
	\end{figure}
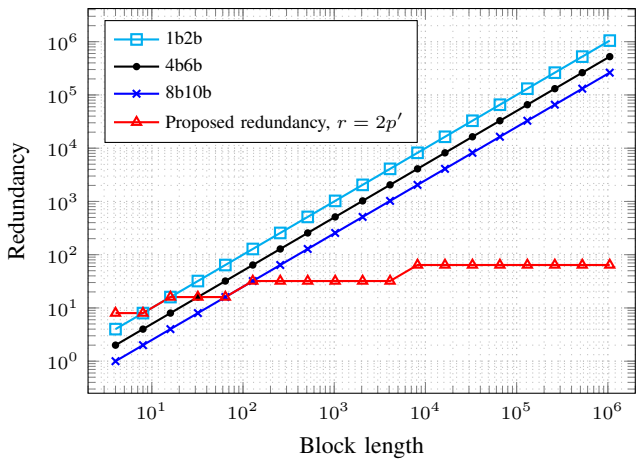

	The efficient communication in VLC relies on limiting the LED power dissipation. This is translated into reducing the redundancy of additional bits while increasing the data transmission rate, which is one of the main achievements of introducing Knuth's algorithm instead of using traditional RLL codes.
	Fig.~\ref{red} presents a comparison of the redundancy performance between Manchester codes, also known as 1b2b, 4b6b, 8b10b, and the proposed one. The redundancy $r$ of our scheme is $r=2 p'$.
	The 1b2b, 4b6b, and 8b10b RLL schemes have redundancies of $N$, $N/2$, and $N/4$ respectively. As shown in Fig.~\ref{red}, the proposed scheme redundancy grows logarithmically whereas all RLL coding schemes have a linear progression which is proportionally inefficient as $N$ increases.
	
	\subsection{RLL Analysis}\label{s3.3}
	In this subsection, we study the run-length characteristic due to the concatenation of polar codes and Knuth's algorithm.
	
	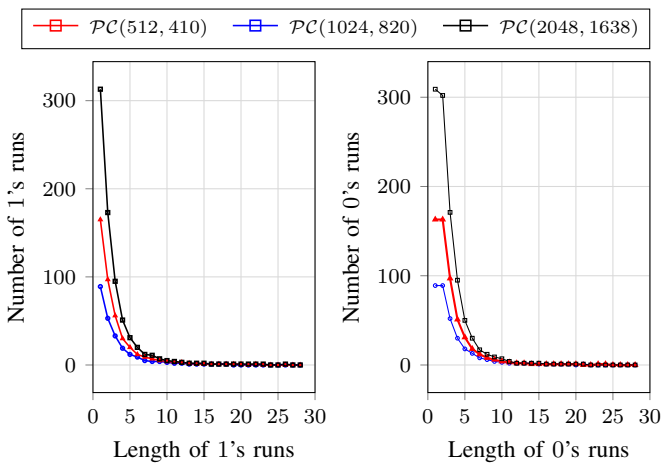
\begin{figure}[t]
		\centering
		\input{img/group_rll.tikz}
		\caption{Distribution of RLLs at $R=0.8$.}
		\label{rll1}
	\end{figure}
	
	Fig.~\ref{rll1} shows the distribution of run-lengths for polar codes of length $512$, $1024$, and $2048$, at a code rate of $0.8$. The information word was generated through uniform distribution, then polar encoded and balanced via Knuth's algorithm with the prefix appended at the transmitted codeword $\bfs$. The numbers of runs versus their lengths were statistically studied for the runs of $1$'s and $0$'s within $\bfs$.
	
	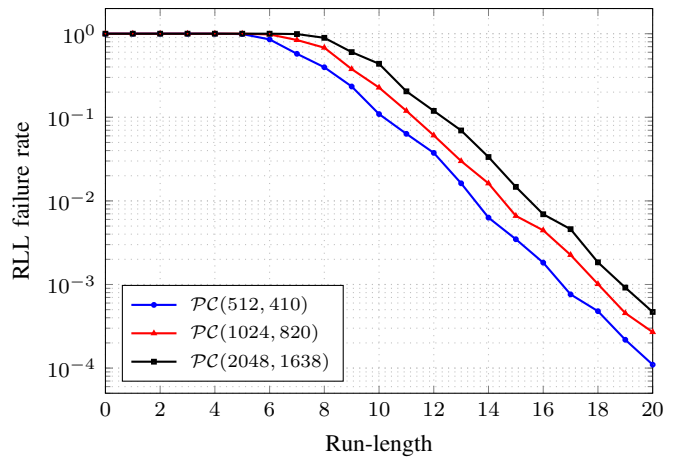
\begin{figure}[t]
		\centering
		\input{img/rllAnalysis.tikz}
		\caption{Run-length performance at $R=0.8$.}
		\label{rll2}
	\end{figure}
	
	For each polar code of lengths $512$, $1024$, and $2048$, $10000$ tests were run as in \cite{fang2017}. About $90\%$ in each generated codeword have a run-length of $l<8$. Nevertheless, the highest run of $l=28$ was recorded at $\mathcal{PC}(2048,1638)$ which occurs once out of $2048\times 10000$ processed bits.
	
	Furthermore, the RLL failure rate in terms of the run-lengths is recorded in Fig.~\ref{rll2} for the same polar codes as above. We observed a decreasing RLL failure rate as run-lengths increase, which are proportional with polar codes of large lengths. This is in correlation with the results of Fig.~\ref{rll1} showing that the number of runs  decays with increasing run-lengths. We approximated the RLL failure rate for $l=28$, which is the highest RLL according to Fig.~\ref{rll1}, to around $10^{-7}$ based on the results of Fig.~\ref{rll2}. 
	
	Considering the lowest optical clock rate of $200$ kHz in VLC systems, the corresponding time period equals $28\times 1/200000=0.14$~ms which is equivalent to a switching frequency of $1/0.14 \mbox{ ms} = 7143$~Hz. The maximum flickering time period of the proposed scheme is very far less than the standard one of $5$~ms and the switching frequency of $7143$~Hz is much higher than the eye-safe one of $200$~Hz.
	
	In accordance with the VLC standard \cite{ieee2011st}, a $15$~MHz LED bandwidth is often used. Therefore, the proposed scheme results in a switching period of $28/(15\times 10^6)\approx 1.87$~$\mu$s (switching frequency of $1/(1.87\times 10^{-6})\approx 535.7$~MHz) when used in the VLC standard.
	
	\subsection{Error Correction Performance}\label{s3.4}
	\subsubsection{Theoretical FER}
	Since the coding scheme relies on the concatenation of two FEC codes, a careful design of the inner FEC code has to be considered. Indeed, if an error occurs during the prefix decoding, regardless of the outer code decoding, the resulting word will be wrong. However, via the theoretical FER prediction under SC decoding, it is possible to obtain a good trade-off between the code rate of the inner code and the decoding performance.
	
	
	
	
	The theoretical FER of the proposed scheme -- if polar codes and SC algorithm are considered -- can be derived as:
	\begin{equation}
	FER = \Big(1-\prod_{i \in \mathcal{I}_2}(1-\bm{Q}_{i})\Big)+\prod_{i \in \mathcal{I}_2}(1-\bm{Q}_{i})\cdot \Big(1-\prod_{i \in \mathcal{I}_1}(1-\bm{Q}_{i})\Big),
	\label{eqmain}
	\end{equation}    
	where $\mathcal{I}_1$ and $\mathcal{I}_2$ are the information sets respectively for the main polar code and for the polar code protecting the prefix.
	
	\begin{figure}[t]
		\centering
		\input{img/FER.tikz}
		\caption{Simulated versus theoretical FER values.}
		\label{fer}
	\end{figure}
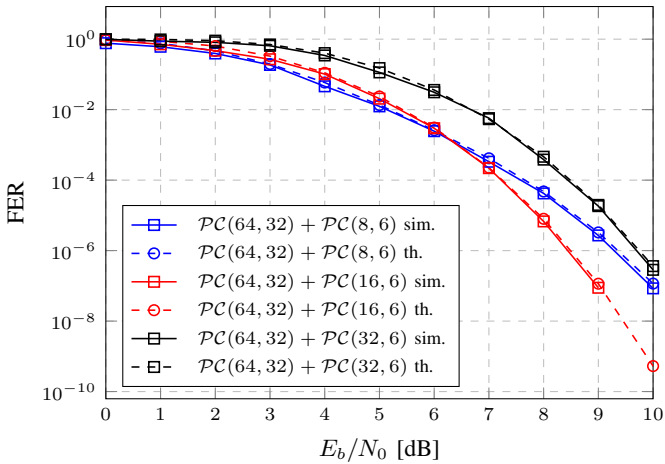
	
	The proposed scheme was implemented and compared with theoretical values for FER as presented in Fig.~\ref{fer} for a $\mathcal{PC}(64,32)$ with a prefix of length $6$ encoded with $\mathcal{PC}(8,6)$, $\mathcal{PC}(16,6)$, and $\mathcal{PC}(32,6)$. The simulated curves (sim.) were validated with a number of $100$ frame errors at $E_b/N_0$ value whereas the theoretical (th.) ones were derived through \eqref{eqmain}. It can be seen that the simulated and the theoretical curves are very close which testify the correctness of our results. One can observe that if the prefix is not sufficiently protected, the scheme cannot reach low error rate (cf. $\mathcal{PC}(8,6)$). On the contrary, using too many redundant bits for the prefix incurs a rate loss, delaying the convergence (cf. $\mathcal{PC}(32,6)$).
	
	\subsubsection{Simulations}
	
	Fig.~\ref{mainRes} shows the BER comparison of the proposed scheme against a polar code with 1b2b encoding and a polar code with 4b6b encoding for transmission rates of $R=0.44$ (square-marked curves) and $R=0.23$ (circle-marked curves). Sizes and rates of the different polar codes have been selected to ensure that the global rate and the number of transmitted bits are the same. For polar code with lengths different than powers of two, the shortening technique in \cite{wang2014} was used to allow length-flexibility, and the soft-input soft-output APP decoding was implemented for the 1b2b and 4b6b schemes. At the rate of $0.44$, the proposed scheme presents a gain of $2.2$~dB and $1.8$~dB over a polar code with 1b2b and a polar code with 4b6b respectively at a BER of $10^{-6}$. For the rate of $0.23$, there is a gain of $0.95$~dB and $2.2$~dB against these schemes at the same BER.
	
	It has been established that RLL codes can achieve capacity for large lengths with high complexity \cite{immink2004}. As a result, there is a trade-off between error correction performance and redundancy for RLL codes. For long codes, the performance of 1b2b is better than that of 4b6b. RLL codes have a high-complexity decoding and are memory consuming in VLC. Therefore, the proposed coding scheme provides an efficient alternative scheme for VLC.

\begin{figure}[t]
		\centering
		\input{img/rate75.tikz}
		\caption{Proposed scheme at $50\%$ dimming ratio and transmission rate of $0.75$.}	
		\label{rate75}
	\end{figure}
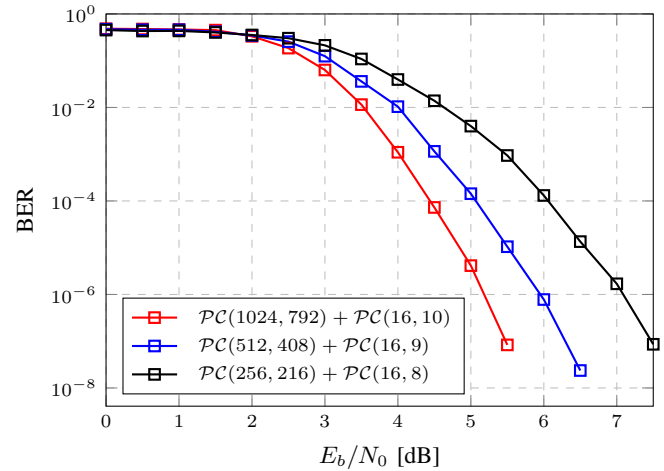

	
	Another limitation of RLL-based techniques is the inherent low transmission rates. For 1b2b and 4b6b codes, the transmission rates are at most $0.5$ and $0.67$ respectively. Fig.~\ref{rate75} presents the performance of the proposed scheme for polar codes $\mathcal{PC}(1024,792)+\mathcal{PC}(16,10)$, $\mathcal{PC}(512,408)+\mathcal{PC}(16,9)$, and $\mathcal{PC}(256,216)+\mathcal{PC}(16,8)$, with a transmission rate of $75\%$. It can be observed that the proposed scheme allows flexibility in transmission rates while maintaining the error correction performance and guaranteeing a flicker-free communication.
	
	\begin{figure}[t]
		\centering
		\input{img/group.tikz}
		\caption{Comparison of various schemes at $50\%$ dimming ratio.}
		\label{mainRes}
	\end{figure}
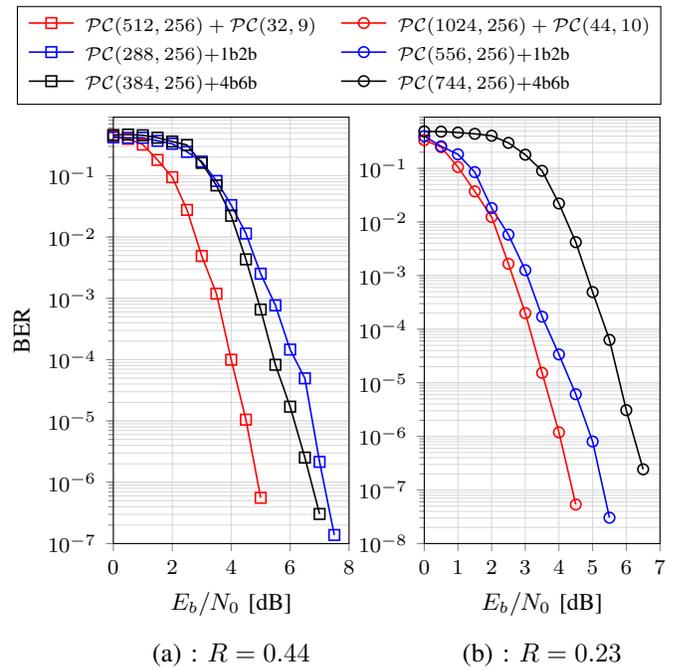

	\subsection{Computational complexity analysis}
	
	\begin{table}[h]
		\centering
		\caption{Number of operations required for decoding RLL codes and the proposed scheme for different rates with $K = 256$.}
		\begin{tabular}{@{}lcc@{}}
			\toprule
			& $R = 1/2$ & $R = 1/4$ \\ 
			\midrule
			1b2b     & $288$     & $556$              \\
			4b6b     & $15744$   & $30504$            \\
			Proposed & $704$     & $1455$             \\ 
			\bottomrule
		\end{tabular}
		\label{tab:cplxity}
	\end{table}
	
	This section aims to evaluate the computational complexity of the proposed scheme, compared with the state-of-the-art balancing scheme. In the following, we assume that the main FEC requires soft information and only the decoding complexity of the balancing operation is considered.
	
	For decoding the 1b2b scheme, $N$ subtractions have to be performed. The APP decoding of each 4b6b block requires two steps. First, probabilities for each possible codeword have to be computed which is performed through $16$ different multiplications of $6$ values, each one obtained by an exponentiation. Second, the marginal probability for each bit is calculated by computing the logarithm of the ratio between the sum of the probabilities of the eight codewords, considering the current bit is either $0$ or $1$. For decoding the proposed scheme, three steps have to be considered. First, the Manchester code is decoded which requires $p'$ subtractions. Second, the polar code protecting the prefix is decoded, involving $M\log_2{M}$ operations, with $M = 2^{\lceil \log_2{p'} \rceil}$. Third, the sign of up to $N$ logarithmic LR (LLR) values have to be inverted which is obtained via $N$ different comparison operations.
	
	Table~\ref{tab:cplxity} summarizes the number of operations for the different methods with $K = 256$ at two distinct rates. All the 2-input operations (addition, multiplication, maximum, exponentiation, logarithmic function, etc.) are considered equivalent. This statement benefits to the 4b6b scheme, which is the only one relying on exponentiation and logarithmic operations. However, the 4b6b exhibits the biggest number of operations, making it impractical in a soft-input soft-output decoding context. The 1b2b scheme has the smallest number of operations. Nonetheless, only code rates below $0.5$ can be achieved. Thus, even with an efficient decoding, the 1b2b scheme highly reduces the achievable spectral efficiency. Finally, the proposed scheme requires less than three times the number of operations required for the 1b2b scheme in the presented cases, making the proposed scheme practical.
	
	\section{Conclusion}\label{s5}
	An efficient method is presented for generating flicker-free codes to mitigate the light flickering in VLC. Results show that the proposed scheme outperforms most state-of-the-art schemes in terms of error correction performance and implementation complexity. The proposed scheme does not use lookup tables which are memory consuming with a complex decoding and low transmission rate. Furthermore, the proposed system is very flexible and can be adapted to other FEC codes for providing flicker-free codes at high transmission rates. These advantages make the proposed scheme suitable for numerous communications applications where balanced and/or flicker-free constraint for the underlying FEC is required.
	
	Future works include extending the proposed scheme to perform efficient dimming in VLC while maintaining flicker-free property, investigating a further compression of the overall prefix for saving power, and improving the error correction performance using sophisticated decoders such as SC list and cyclic redundancy check (CRC) aided SC list.
	\vspace*{.7em}
	\section*{Acknowledgments}
	S.~A.~Hashemi is supported by a Postdoctoral Fellowship from the Natural Sciences and Engineering Research Council of Canada (NSERC).
	\vspace*{.5em}
	\bibliographystyle{IEEEtran}
	\bibliography{IEEEfull,references}
\end{document}

%% file: img/rllResults.tikz
\begin{tikzpicture}[spy using outlines = {rectangle, magnification=2, connect spies}]
\pgfplotsset{	
	label style = {font=\fontsize{9pt}{7.2}\selectfont},
	tick label style = {font=\fontsize{7pt}{7.2}\selectfont}
}

\begin{axis}[
scale = 1,
ymode=log,
xlabel={Run-length}, 
ylabel={RLL failure rate}, 
xtick={0,2,4,6,8,10,12,14,16,18,20,22,24},
grid=both,
ymin=1e-5,
xmin = 0,
xmax = 24,
ymajorgrids=true,
xmajorgrids=true,
grid style=dotted,
width=\columnwidth, height=6.7cm,
mark size=3,
    legend style={
     anchor={south east},
      cells={anchor=west},
      column sep= 1mm,
      font=\fontsize{7pt}{7.2}\selectfont,
},
legend pos=south west
]

\addplot[
color=blue,
mark=square,
thick,
mark size=0.7,
]
table[x=LR, y=Turbo, col sep=comma]{img/RLLS.csv};
\addlegendentry{Turbo code $(1024,512)$}


\addplot[
color=black,
mark=square,
thick,
mark size=0.6,
]
table[x=LR, y=LDPC, col sep=comma]{img/RLLS.csv};
\addlegendentry{LPDC code $(1024,512)$}

  
  \addplot[
  color=red,
  mark=square,
  thick,
  mark size=0.6,
  ]
  table[x=LR, y=Polar_GA, col sep=comma]{img/RLLS.csv};
  \addlegendentry{$\mathcal{PC}(1024,512)$}
  
\end{axis}
\end{tikzpicture}

%% file: img/figure.tikz
\begin{tikzpicture}[scale=1, spy using outlines = {rectangle, magnification=3, connect spies}]
  \pgfplotsset{	
    label style = {font=\fontsize{9pt}{7.2}\selectfont},
    tick label style = {font=\fontsize{7pt}{7.2}\selectfont}
  }

\begin{axis}[
	scale = 1,
    ymode=log,
    xmode=log,
    xlabel={Block length}, 
    ylabel={Redundancy}, 
    xmin =2,
    xmax =2e6, 
    grid=both,
    ymajorgrids=true,
    xmajorgrids=true,
    grid style=dotted,
    width=\columnwidth, height=6.7cm,
   legend style={
     anchor={south east},
     cells={anchor=west}, 
     font=\fontsize{7pt}{7.2}\selectfont,,
   },
    legend pos=north west
]

\addplot[color=cyan, mark=square, mark size=2,thick]
table[x=X, y=M, col sep=comma]{img/Book1.csv};
\addlegendentry{1b2b}

\addplot[color=black, mark=*, mark size=1, thick, ]
table[x=X, y=4B, col sep=comma]{img/Book1.csv};
\addlegendentry{4b6b}

\addplot[color=blue, mark=x, mark size=2, thick]
table[x=X, y=8B, col sep=comma]{img/Book1.csv};;
\addlegendentry{8b10b}

\addplot[color=red, mark=triangle, mark size=2, thick]
table[x=X, y=K1, col sep=comma]{img/Book1.csv};
\addlegendentry{Proposed redundancy, $r=2p'$ }

\end{axis}

\end{tikzpicture}

%% file: img/group_rll.tikz
\begin{tikzpicture}
    \pgfplotsset{   
        label style = {font=\fontsize{7pt}{7}\selectfont},
        tick label style = {font=\fontsize{7pt}{7}\selectfont}
    }
    \begin{groupplot}[group style={group name=fer_group, group size= 2 by 1, horizontal sep=1.5cm, vertical sep=0cm}, 
                      footnotesize,
                      height=0.33\textwidth,  width=.25\textwidth,
                      tick align=outside, tickpos=left,
                      grid=both, grid style={gray!30},
                      xmin = 0,
                      xmax = 30,
                      ytick = {0, 100, 200, 300},
             ]

            \nextgroupplot[xlabel={Length of $1$'s runs}, ylabel={Number of $1$'s runs},]
            \addplot[
                color=blue,
                mark=o,
                semithick,
                mark size=0.7,
                ]
                table[x=L, y=KR1, col sep=comma]{img/run1.csv};\label{plots:plot1}
                
            \addplot[
                color=red,
                mark=triangle,
                semithick,
                mark size=0.7,
                ]
                table[x=L, y=KR12, col sep=comma]{img/run1.csv};\label{plots:plot2}
                
            \addplot[
                color=black,
                mark=square,
                semithick,
                mark size=0.7,
                ]
                table[x=L, y=KR13, col sep=comma]{img/run1.csv};\label{plots:plot3}
                \coordinate (top) at (rel axis cs:0,1);

            \nextgroupplot[xlabel={Length of $0$'s runs}, ylabel={Number of $0$'s runs},]

            \addplot[
color=blue,
mark=o,
mark size=0.7,
]
table[x=L, y=KR0, col sep=comma]{img/run1.csv};

\addplot[
color=red,
mark=triangle,
thick,
mark size=0.7,
]
table[x=L, y=KR02, col sep=comma]{img/run1.csv};

\addplot[
color=black,
mark=square,
mark size=0.7,
]
table[x=L, y=KR03, col sep=comma]{img/run1.csv};

\coordinate (bot) at (rel axis cs:1,0);
\end{groupplot}

\path (top|-current bounding box.north) -- 
      coordinate(legendpos)
      (bot|-current bounding box.north);
    \matrix[
        matrix of nodes,
        anchor=south,
        draw,
        inner sep=0.2em,
        draw
      ]at([yshift=1ex, xshift=-3ex]legendpos)
      {
        \ref{plots:plot1}& \fontsize{7pt}{7.2}\selectfont $\mathcal{PC}(512,410)$  &[5pt]
        \ref{plots:plot2}& \fontsize{7pt}{7.2}\selectfont $\mathcal{PC}(1024,820)$ &[5pt]
        \ref{plots:plot3}& \fontsize{7pt}{7.2}\selectfont $\mathcal{PC}(2048,1638)$ \\        };

\end{tikzpicture}

%% file: img/rllAnalysis.tikz
\begin{tikzpicture}[spy using outlines = {rectangle, magnification=2, connect spies}]
\pgfplotsset{	
	label style = {font=\fontsize{9pt}{7.2}\selectfont},
	tick label style = {font=\fontsize{7pt}{7.2}\selectfont}
}

\begin{axis}[
footnotesize,
scale = 1,
ymode=log,
xlabel={Run-length}, 
ylabel={RLL failure rate}, 
xtick={0,2,4,6,8,10,12,14,16,18,20},
grid=both,
ymin=5e-5,
ymax = 2,
xmin = 0,
xmax = 20,
ymajorgrids=true,
xmajorgrids=true,
grid style=dotted,
width=\columnwidth, height=6.7cm,
mark size=3,
    legend style={
     anchor={south east},
      cells={anchor=west},
      column sep= 1mm,
      font=\fontsize{7pt}{7.2}\selectfont,
},
legend pos=south west
]

\addplot[
color=blue,
mark=o,
thick,
mark size=0.7,
]
table[x=R, y=F1, col sep=comma]{img/RLLS1.csv};
\addlegendentry{$\mathcal{PC}(512,410)$}

\addplot[
color=red,
mark=triangle,
thick,
mark size=0.7,
]
table[x=R, y=F2, col sep=comma]{img/RLLS1.csv};
\addlegendentry{$\mathcal{PC}(1024,820)$}

\addplot[
color=black,
mark=square,
thick,
mark size=0.7,
]
table[x=R, y=F3, col sep=comma]{img/RLLS1.csv};
\addlegendentry{$\mathcal{PC}(2048,1638)$}
  
\end{axis}
\end{tikzpicture}

%% file: img/FER.tikz
\begin{tikzpicture}[spy using outlines = {rectangle, magnification=2.5, connect spies}]
	\pgfplotsset{	
		label style = {font=\fontsize{9pt}{7}\selectfont},
		tick label style = {font=\fontsize{7pt}{7}\selectfont}
	}
	
	\begin{axis}[
	scale = 1,
	ymode=log,
	xlabel={$E_b/N_0$ [\text{dB}]}, 
	ylabel={FER}, 
	xtick={0,1,2,3,4,5,6,7,8,9,10,11,12,13,14,15,16},
	xmax = 10,
	xmin = 0,
	grid=both,
	legend cell align={left},
	ymajorgrids=true,
	xmajorgrids=true,
	grid style=dashed,
    width=\columnwidth, height=6.8cm,
	mark size=2,
	legend style={
	anchor={south east},
	cells={anchor=west},
	column sep= 2mm,
	font=\fontsize{7pt}{7.2}\selectfont,
},
legend pos=south west
	]
\addplot[
	color=blue,
	mark=square,
	mark options={solid},
	mark size=2,
	semithick,
	]
	table {
0 7.69E-01
1 6.10E-01
2 3.93E-01
3 1.86E-01
4 4.59E-02
5 1.24E-02
6 2.45E-03
7 3.33E-04
8 4.20E-05
9 2.70E-06
10 8.49e-08
	};
	\addlegendentry{$\mathcal{PC}(64,32)+\mathcal{PC}(8,6)$ sim.}
	
	\addplot[
	color=blue,
	mark=o,
	dashed,
	mark options={solid},
	mark size=2,
	semithick,
	]
	table {
0 9.33E-01
1 7.60E-01
2 4.65E-01
3 1.97E-01
4 5.87E-02
5 1.35E-02
6 2.61E-03
7 4.16E-04
8 4.76E-05
9 3.28E-06
10 1.16E-07	
	};
	\addlegendentry{$\mathcal{PC}(64,32)+\mathcal{PC}(8,6)$ th.}
	
	\addplot[
	color=red,
	mark=square,
	mark options={solid},
	mark size=2,
	semithick,
	]
	table {
		0 9.43E-01
1 7.15E-01
2 4.72E-01
3 2.69E-01
4 1.02E-01
5 2.05E-02
6 2.97E-03
7 2.19E-04
8 6.74E-06	
9 8.84e-08
	};
	\addlegendentry{$\mathcal{PC}(64,32)+\mathcal{PC}(16,6)$ sim.}
	
	\addplot[
	color=red,
	mark=o,
	dashed,
	mark options={solid},
	mark size=2,
	semithick,
	]
	table {
0 9.71E-01
1 8.73E-01
2 6.38E-01
3 3.30E-01
4 1.11E-01
5 2.36E-02
6 3.09E-03
7 2.26E-04
8 7.99E-06
9 1.15E-07
10 5.29E-10
	
	};
	\addlegendentry{$\mathcal{PC}(64,32)+\mathcal{PC}(16,6)$ th.}
	
	\addplot[
	color=black,
	mark=square,
	mark options={solid},
	mark size=2,
	semithick,
	]
	table {
		0 9.90E-01
1 8.77E-01
2 8.13E-01
3 6.49E-01
4 3.45E-01
5 1.14E-01
6 3.11E-02
7 5.74E-03
8 3.82E-04
9 1.83E-05
10 2.85e-07
	};
	\addlegendentry{$\mathcal{PC}(64,32)+\mathcal{PC}(32,6)$ sim.}

	\addplot[
	color=black,
	mark=square,
	dashed,
	mark options={solid},
	mark size=2,
	semithick,
	]
	table {
		0 9.96E-01
1 9.79E-01
2 9.06E-01
3 7.06E-01
4 4.02E-01
5 1.51E-01
6 3.62E-02
7 5.37E-03
8 4.61E-04
9 1.98E-05
10 3.63E-07	
	};
	\addlegendentry{$\mathcal{PC}(64,32)+\mathcal{PC}(32,6)$ th.}
	\end{axis}
	\end{tikzpicture}

%% file: img/group.tikz
\begin{tikzpicture}
    \pgfplotsset{   
        label style = {font=\fontsize{7pt}{7}\selectfont},
        tick label style = {font=\fontsize{7pt}{7}\selectfont}
    }
    \begin{groupplot}[group style={group name=fer_group, group size= 2 by 1, horizontal sep=1cm, vertical sep=0cm}, 
                      footnotesize,
                      height=0.4\textwidth,  width=.26\textwidth,
                      xlabel= $E_b/N_0 \text{~[dB]}$,
                      ymode=log,
                      ymax = .9,
                      tick align=outside, tickpos=left,
                      grid=both, grid style={gray!30},
             ]

    \nextgroupplot[xmin = 0, xmax = 8, ylabel= BER, ymin = 1e-7]
    \addplot[
    color=red,
    mark=square,
    semithick,
    mark options={solid},
    mark size=2,
    ]
    table {
        0 4.43E-01
        .5 4.00E-01
        1 3.22E-01
        1.5 1.81E-01
        2 9.46E-02
        2.5 2.77E-02
        3 4.91E-03
        3.5 1.19E-03
        4 9.98E-05
        4.5 1.05E-05
        5 5.58E-07
        
    };\label{plots:plot1}
    
    \addplot[
    color=blue,
    mark=square,
    semithick,
    mark size=2,
    ]
    table {
        0 4.23E-01
        .5 4.15E-01
        1 4.16E-01
        1.5 3.71E-01
        2 3.32E-01
        2.5 2.45E-01
        3 1.70E-01
        3.5 8.22E-02
        4 3.32E-02
        4.5 1.14E-02
        5 2.52E-03
        5.5 7.70E-04
        6 1.46E-04
        6.5 4.96E-05
        7 2.15E-06
        7.5 1.38E-07    
    };\label{plots:plot2}

    \addplot[
    color=black,
    mark=square,
    semithick,
    mark size=2,
    ]
    table {
        0 4.61E-01
        .5 4.68E-01
        1 4.57E-01
        1.5 4.17E-01
        2 3.61E-01
        2.5 3.17E-01
        3 1.62E-01
        3.5 6.94E-02
        4 2.22E-02
        4.5 4.32E-03
        5 6.57E-04
        5.5 8.24E-05
        6 1.71E-05
        6.5 2.53E-06
        7 3.04e-07
        
    };\label{plots:plot3}
    \coordinate (top) at (rel axis cs:0,1);

    \nextgroupplot[xmin = 0, xmax = 7, ymin = 1e-8]
    \addplot[
    color=red,
    mark=o,
    semithick,
    mark options={solid},
    mark size=2,
    ]
    table {
        0 3.36E-01
        .5 2.49E-01
        1 1.06E-01
        1.5 3.72E-02
        2 1.24E-02
        2.5 1.66E-03
        3 1.99E-04
        3.5 1.53E-05
        4 1.19E-06
        4.5 5.35e-08
            
    };\label{plots:plot4}
    
    \addplot[
    color=blue,
    mark=o,
    semithick,
    mark size=2,
    ]
    table {
        0 3.99E-01
        .5 2.58E-01
        1 1.83E-01
        1.5 8.52E-02
        2 1.82E-02
        2.5 5.80E-03
        3 1.26E-03
        3.5 1.71E-04
        4 3.38E-05
        4.5 6.11E-06
        5 8.02E-07
        5.5 3.04E-08
        
    };\label{plots:plot5}
    
    \addplot[
    color=black,
    mark=o,
    semithick,
    mark size=2,
    ]
    table {
        0 4.89E-01
        .5 4.84E-01
        1 4.66E-01
        1.5 4.41E-01
        2 4.06E-01
        2.5 2.99E-01
        3 1.80E-01
        3.5 8.99E-02
        4 2.21E-02
        4.5 4.20E-03
        5 4.92E-04
        5.5 6.31E-05
        6 3.08E-06
        6.5 2.43e-07
    };\label{plots:plot6}
    \coordinate (bot) at (rel axis cs:1,0);
     \end{groupplot}

    \node[below = 1.2cm of fer_group c1r1.south] {(a) : $R = 0.44$};
    \node[below = 1.2cm of fer_group c2r1.south] {(b) : $R = 0.23$};

     \path (top|-current bounding box.north) -- 
      coordinate(legendpos)
      (bot|-current bounding box.north);
    \matrix[
        matrix of nodes,
        anchor=south,
        draw,
        inner sep=0.2em,
        draw,
        column 2/.style={anchor=base west},
        column 4/.style={anchor=base west}
      ]at([yshift=1ex, xshift=-4ex]legendpos)
      {
        \ref{plots:plot1}& \fontsize{7pt}{7.2}\selectfont $\mathcal{PC}(512,256)+\mathcal{PC}(32,9) $&[5pt]
        \ref{plots:plot4}& \fontsize{7pt}{7.2}\selectfont $\mathcal{PC}(1024,256)+\mathcal{PC}(44,10)$ \\
        \ref{plots:plot2}& \fontsize{7pt}{7.2}\selectfont $\mathcal{PC}(288,256)+$1b2b &[5pt]
        \ref{plots:plot5}& \fontsize{7pt}{7.2}\selectfont $\mathcal{PC}(556,256)+$1b2b \\
        \ref{plots:plot3}& \fontsize{7pt}{7.2}\selectfont $\mathcal{PC}(384,256)+$4b6b &[5pt]
        \ref{plots:plot6}& \fontsize{7pt}{7.2}\selectfont $\mathcal{PC}(744,256)+$4b6b \\        };
  \end{tikzpicture}

%% file: img/rate75.tikz
\begin{tikzpicture}[spy using outlines = {rectangle, magnification=2.5, connect spies}]
	\pgfplotsset{	
		label style = {font=\fontsize{9pt}{7}\selectfont},
		tick label style = {font=\fontsize{7pt}{7}\selectfont}
	}
	
	\begin{axis}[
	scale = 1,
	ymode=log,
	xlabel={$E_b/N_0$ [\text{dB}]}, 
	ylabel={BER}, 
	xtick={0,1,2,3,4,5,6,7},
	ymax = 1,
	xmin=0,
	xmax = 7.5,
	grid=both,
	legend cell align={left},
	ymajorgrids=true,
	xmajorgrids=true,
	grid style=dashed,
	width=\columnwidth, height=6.8cm,
	mark size=2,
		legend style={
		anchor={south east},
		cells={anchor=west},
		column sep= 2mm,
		font=\fontsize{7pt}{7.2}\selectfont,
	},
	legend pos=south west
	]
	
	\addplot[
	color=red,
	mark=square,
	thick,
	mark options={solid},
	mark size=2,
	]
	table {
		0    4.83e-01
		.5   4.76e-01
		1    4.66e-01
		1.5  4.51e-01
		2    3.33e-01
		2.5  1.86e-01
		3    6.33e-02
		3.5  1.15e-02
		4    1.10e-03
		4.5  7.26e-05
		5    4.14e-06
		5.5  8.34e-08
	};
	\addlegendentry{$\mathcal{PC}(1024,792)+\mathcal{PC}(16,10)$}
	
	\addplot[
	color=blue,
	mark=square,
	thick,
	mark size=2,
	]
	table {
		0    4.72e-01
		0.5  4.59e-01
		1    4.59e-01
		1.5  4.16e-01
		2    3.47e-01
		2.5  2.56e-01
		3    1.25e-01
		3.5  3.61e-02
		4    1.04e-02
		4.5  1.15e-03
		5    1.43e-04
		5.5  1.05e-05
		6    7.79e-07
		6.5  2.36e-08 	
	};
	\addlegendentry{$\mathcal{PC}(512,408)+\mathcal{PC}(16,9)$}

	\addplot[
	color=black,
	mark=square,
	thick,
	mark size=2,
	]
	table {
		0    4.52e-01
		0.5  4.31e-01
		1    4.34e-01
		1.5  4.02e-01
		2    3.55e-01
		2.5  3.03e-01
		3    2.13e-01
		3.5  1.09e-01
		4    3.96e-02
		4.5  1.39e-02
		5    3.99e-03
		5.5  9.38e-04
		6    1.31e-04
		6.5  1.35e-05
		7    1.69e-06
		7.5  8.59e-08
		
	};
	\addlegendentry{$\mathcal{PC}(256,216)+\mathcal{PC}(16,8)$}
	
	\end{axis}
	\end{tikzpicture}